\documentclass{PoS}

\title{Episodic Activity in Active Galactic Nuclei}

\ShortTitle{Episodic Activity in Active Galactic Nuclei}

\author{\speaker{D. J. Saikia} \\ 
        NCRA, TIFR, Pune, India \\
        E-mail: \email{djs@ncra.tifr.res.in}}

\author{M. Jamrozy \\
        Uniwersytet Jagiello\'nski, Krak\'ow, Poland\\
        E-mail: \email{jamrozy@oa.uj.edu.pl}}

\author{C. Konar \\
        ASIAA, Taiwan \\
        E-mail: \email{chiranjib.konar@gmail.com}}

\author{S. Nandi \\
        ARIES, Nainital \\
        E-mail: \email{sumana1981@gmail.com}}

\abstract{There is increasing evidence to suggest that AGN activity
          may be episodic, with a wide range of possible time scales.
          Radio galaxies exhibit the most striking examples of episodic
          activity, with two or three distinct pairs of lobes on opposite
          sides of the active nucleus. Radio emission from earlier cycles
          of activity are expected to have steep radio spectra due to 
          radiative losses, and hence be detected more easily
          at low radio frequencies. Inverse-Compton scattered cosmic 
          microwave background radiation could in prinicple probe even
          lower Lorentz-factor particles, revealing an older population.
          We illustrate the time scales of episodic activity by considering
          different radio galaxies, discuss the possiblity of episodic activity
          in cluster radio sources, and a possible trend for a  high incidence
          of H{\sc i} absorption in sources with evidence of episodic activity.
          }

\FullConference{25th Texas Symposium on Relativistic Astrophysics - TEXAS 2010\\
		December 06-10, 2010\\
		Heidelberg, Germany}

\begin{document}

\section{Introduction}
It is generally believed that the energy source of active galactic nuclei (AGN) is
accretion of matter onto a supermassive black hole (SMBH) with masses ranging from 
$\sim$10$^6$ to 10$^{10}$ M$_\odot$ \cite{1}. In this paradigm, periodic `feeding' of 
the SMBH could lead to episodic AGN activity. Alternative models for intermittency
are related to successful jet formation due to collimation by MHD outflows from the
accretion disks  {\cite{2,3}.  Marconi 
et al. \cite{4}  consider a scenario is which the black holes in the local Universe have 
grown by mass accretion during AGN phases with the average life time of these phases 
ranging from $\sim$1.5$\times$10$^8$ to 10$^9$ yr. A more extended review of 
recurrent activity in AGN can be found in Saikia \& Jamrozy \cite{5}. 

Amongst galaxies harbouring an AGN, $\sim$10 per cent of objects are radio loud, 
and there have been suggestions that this too may be episodic. For example, the 
possibility
that the essential difference between a radio-loud and a radio-quiet quasar may be the
epoch at which it is being observed has been suggested \cite{6}. 
For the radio loud objects with extended lobes of emission on opposite sides of the
nucleus, the structure and spectra of the lobes contain an imprint of the history of the
source, such as interaction with the environment, radiative losses, reacceleration of
particles and episodic activity of the AGN. One of the very striking examples of 
episodic activity in a radio galaxy is when a new pair of radio lobes is seen in 
addition to the outer lobes from an earlier cycle of activity. Although these have
been christened more recently as `double-double' radio galaxies (DDRGs) \cite{7},
examples of such galaxies have been reported earlier \cite{8,9}. An example of 
a triple-double radio galaxy has also been reported by Brocksopp et al. \cite{42}. 

An interesting way of probing relic emission is via X-ray observations of  
inverse-Compton-scattered Cosmic Microwave Background (ICCMB) radiation,  
which could probe lower Lorentz-factor particles and hence an older population \cite{45}.
Interesting cases of relic emission identified from X-ray observations include
the radio galaxies Cygnus A \cite{10,11}, 3C294 \cite{12} and 4C23.56 \cite{46}.

Deep X-ray and low-frequency radio imaging studies of groups and clusters of
galaxies have also revealed evidence of sources with possibly two or three cycles
of activity. An interesting example is NGC5044 where {\it Chandra} observations 
reveal that the group hosts several radio-quiet cavities, filaments and a semi-circular
cold front. The radio observations reveal possible evidence of multiple cycles of 
activity \cite{13,14}. 

\begin{figure}
\begin{center}
  \hbox{
\centerline{\includegraphics[angle=0, width=.82\textwidth,]{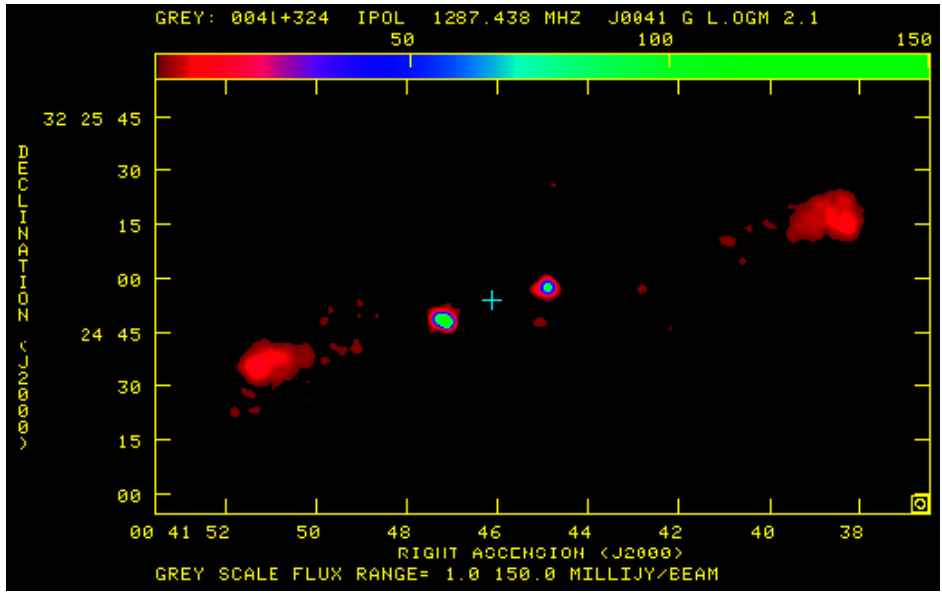}}
       }
\caption{GMRT image of the DDRG J0041+3224 discovered using the Giant Metrewave
Radio Telescope (GMRT) \cite{15}. }
\end{center}
\end{figure}

\section{Examples of DDRGs}
As an example, the radio image of the DDRG J0041+3224 discovered using the 
Giant Metrewave Radio Telescope by Saikia, Konar \& Kulkarni 2006 \cite{15} is
shown in Figure 1. More than about a dozen good
examples of such DDRGs are known. The size of the inner double could be small as
$\sim$14 pc as in the giga-Hertz peaked spectrum (GPS) core of J1247+6723 to several hundred 
kpc as in J1835+6024 \cite{9}. The size of the outer doubles is usually large,
often over a Mpc, although the highly misaligned DDRG 3C293 has an overall size
of only $\sim$190 kpc \cite{16}  and the possibility that the GPS object CTA21 with an overall linear
size of $\sim$0.3 kpc might exhibit evidence of episodic activity has been explored \cite{17}.
However, the number of sources with unambiguous evidence of
episodic activity is still small, and are often associated with large radio galaxies.
This is consistent with the study by Sirothia et al. \cite{18}  where for a sample of
374 sources with a median angular size less than about 10 arcsec, there is no clear
example of a DDRG.

However, there are also selection effects in trying to identify these sources.
The unambiguous examples of DDRGs tend to have an edge-brightened structure 
for the inner double, so that these are distinguishable from knots in the jet
or peaks of emission in the lobe caused by re-acceleration of particles in the lobes. 
The formation of the inner hotspots has been explored recently by Brocksopp et al. \cite{43}.  
For these sources as well, if the time scale of jet interruption is smaller than
the time required for the jet channel to collapse, the jets may not form distinct
pairs of lobes. For an FRI low-luminosity radio galaxy with diffuse lobes of 
emission, a possible way of distingishing different cycles of activity could 
be from spectral-index distributions. A sharp gradient in spectral index may
help demarcate the steeper-spectrum emission from an earlier cycle of activity 
from the more recent emission with a flatter spectrum. The radio structures may
also appear to be different with the low-frequency images detecting more extended
emission from an earlier cycle of activity. However one has to be careful that
these differences are not caused by differences in the resolution or coverage of the
uv plane. Also, the low-luminosity radio sources are often in clusters of galaxies,
and regions of flatter spectral indices may also be caused by turbulence and
re-acceleration of particles as the radio plasma interacts with the intra-cluster
medium. Nevertheless, possible examples of tailed FRI radio galaxies with evidence
of recurrent activity have been identified from radio and X-ray studies of clusters
of galaxies using spectral and structural information. One of the earlier examples
(A2372) was reported by Giacintucci et al. \cite{19}, while a few more possible examples
have been reported recently by Giacintucci et al. \cite{20}. In the case of A2372
Giacintucci et al. found a significant emission gap between the inner jets coincident
with the optical galaxy and the edge of the outer lobes, which they interpreted to 
be due to recurrent activity.  An example of an FRI source
without distinct hot spots, but where at least two cycles of AGN activity are well
accepted from radio structural information is Cen A \cite{21}. In the large radio
galaxy Her A, Gizani \& Leahy \cite{22} have found a somewhat sharp bounday in the 
spectral index distribution in the lobes, which could be interpreted to be due
to different cycles of activity.  

If the DDRGs are associated with only galaxies, it would not be consistent with
the unified scheme where the radio galaxies and quasars are believed to be
intrinsically similar, but differing only in orientation. Since all the reported
DDRGs were associated with galaxies, we tried to identify DDRGs associated with
quasars. From a systematic comparison of FIRST and NVSS images of a large number
of radio selected quasars, and imaging with the GMRT, one  of the most promising
examples of episodic activity in a quasar is 4C02.27 (Figure 2). Diffuse radio
emission, possibly from an earlier cycle, has been imaged with the GMRT, and 
shown to have a steep spectral index of $\sim$1.2 between 610 and 1400 MHz, while
that of the western lobe is $\sim$1.0 \cite{23},.
There is a prominent hot-spot in the western lobe, constraining the time
scale of episodic activity to be $\sim$1.8$-$3.9 Myr, depending on the orientation
of the source \cite{23}. Considering the relatively small number of giant radio
quasars \cite{28}, evidence of episodic activity in 4C02.27 with an overall projected
linear size of $\sim$470 kpc is consistent with the unified scheme for radio galaxies
and quasars \cite{44}. 

\begin{figure}
\begin{center}
\vbox{
  \hbox{
   \includegraphics[angle=0, width=.32\textwidth,]{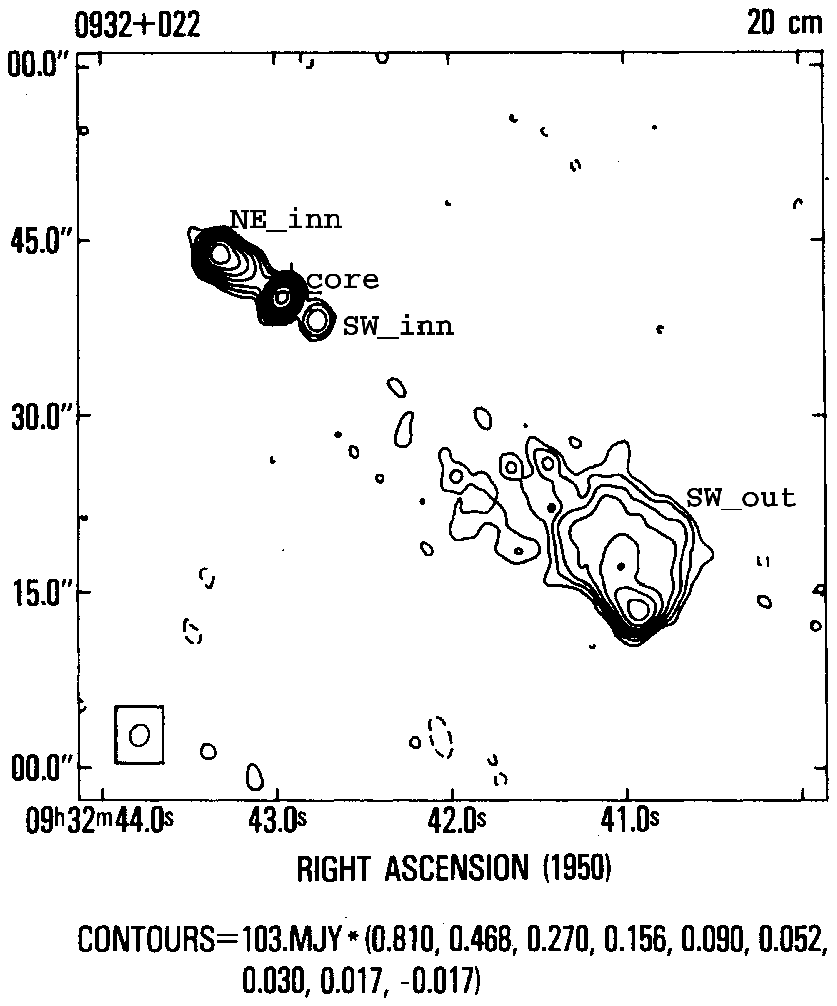}
   \includegraphics[angle=0, width=.50\textwidth,]{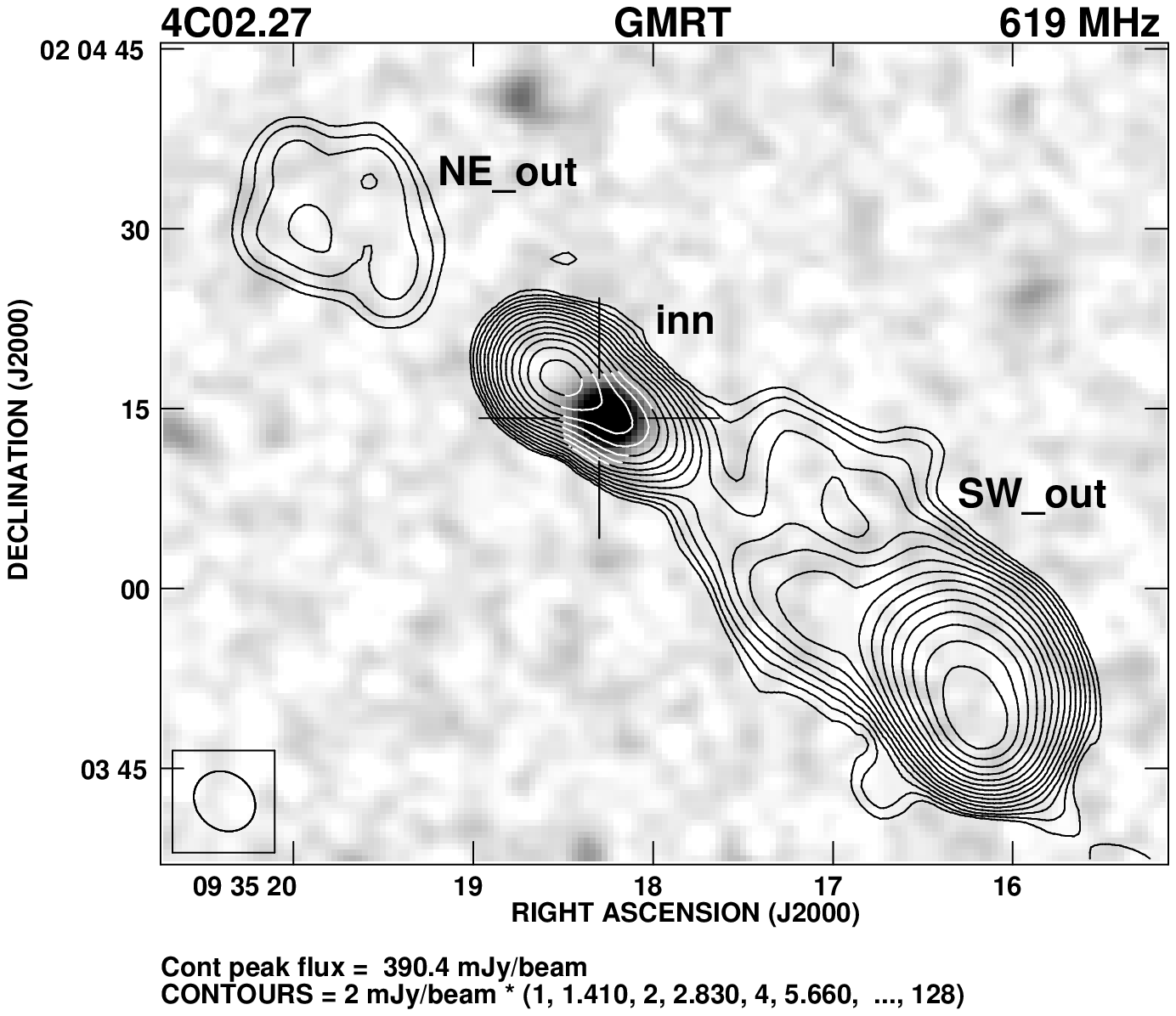}
       }
}
\caption{A Very Large Array image of the double-double radio quasar 4C02.27 at 1413 MHz 
\cite{39} on the left panel and a GMRT image at 619 MHz \cite{23} showing the diffuse
outer lobe on the eastern side. }
\end{center}
\end{figure}

\section{Time scales}
For a sample of 5 DDRGs, Kaiser, Schoenmakers \& R\"ottgering \cite{24} find the
difference in dynamical ages between the outer and inner doubles to be $\approx$10$^8$
yr. Spectral ageing studies of the DDRGs J0840+2949 \cite{25}, 
J1453+3308 \cite{26} and J1548$-$3216 \cite{27},
suggest the difference in spectral ages between the inner and outer pairs to range
from 5$\times$10$^7$ to 10$^8$ yr. However, many of these sources are over a Mpc,
belonging to the class of giant radio sources (e.g. \cite{28}).
Considering smaller sources, the total projected linear size of the inner double of 
of the highly misaligned DDRG 3C293 (J1352+316) is $\sim$4.2 kpc, while that of the outer
double is 190 kpc. The spectral ages of the inner and outer doubles, and the 
existence of a hot-spot suggest that the time scale of interruption of jet activity
is less than $\sim$0.1 Myr \cite{16}. In the case of Cygnus A, with an overall linear 
size of 136 kpc, the 
time scale of interruption of jet activity has been estimated to be $\approx$10$^6$ yr.
The compact radio source, CTA21, whose radio structure has been suggested to show
evidence of recurrent activity, the time scale could be in the range of 
10$^4$-10$^5$ yr \cite{40}. 
The different time scales are being probed using sources of different sizes.

\begin{figure}
\begin{center}
\vbox{
  \hbox{
   \includegraphics[angle=0, width=.32\textwidth,]{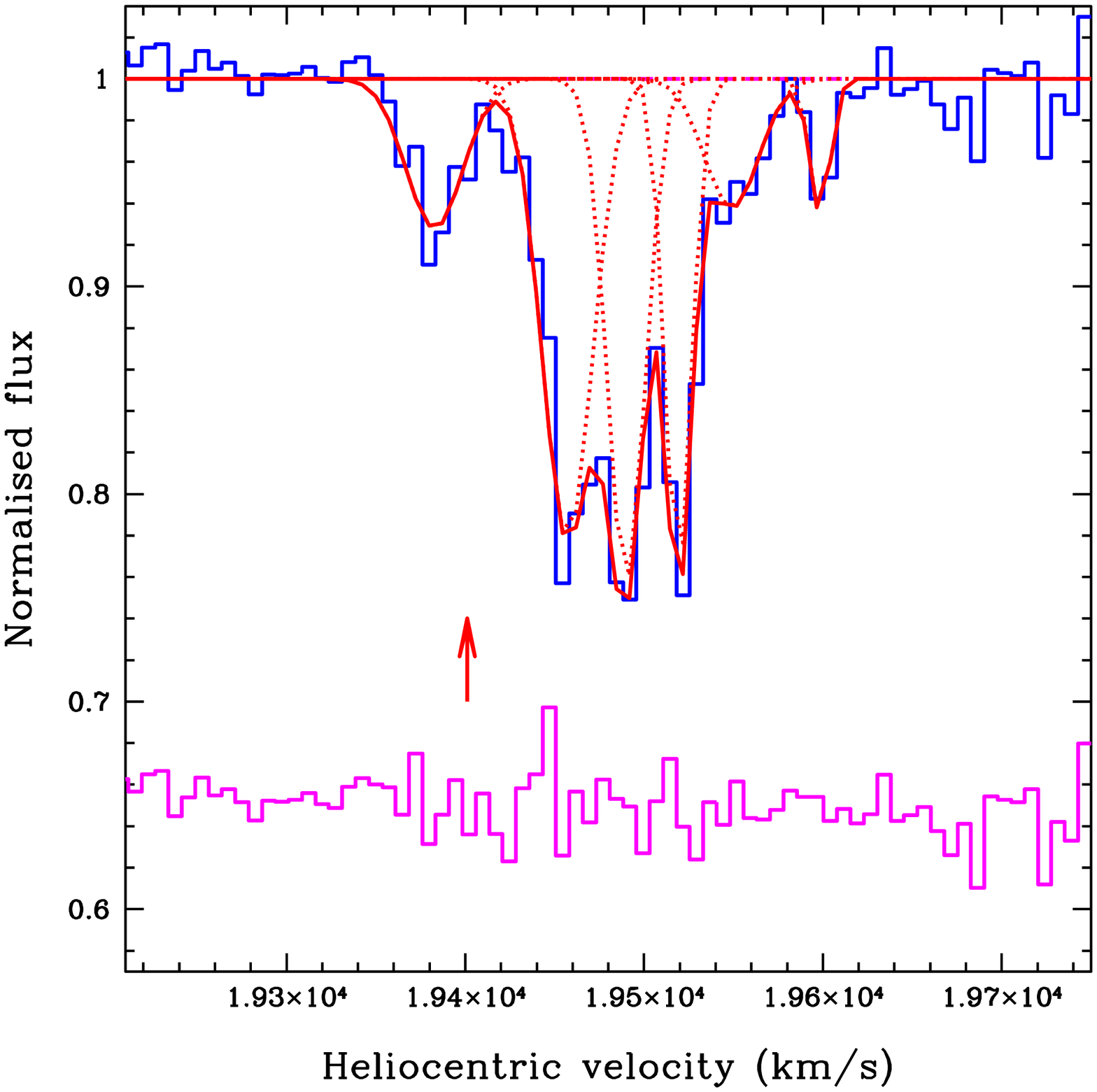}
   \includegraphics[angle=0, width=.32\textwidth,]{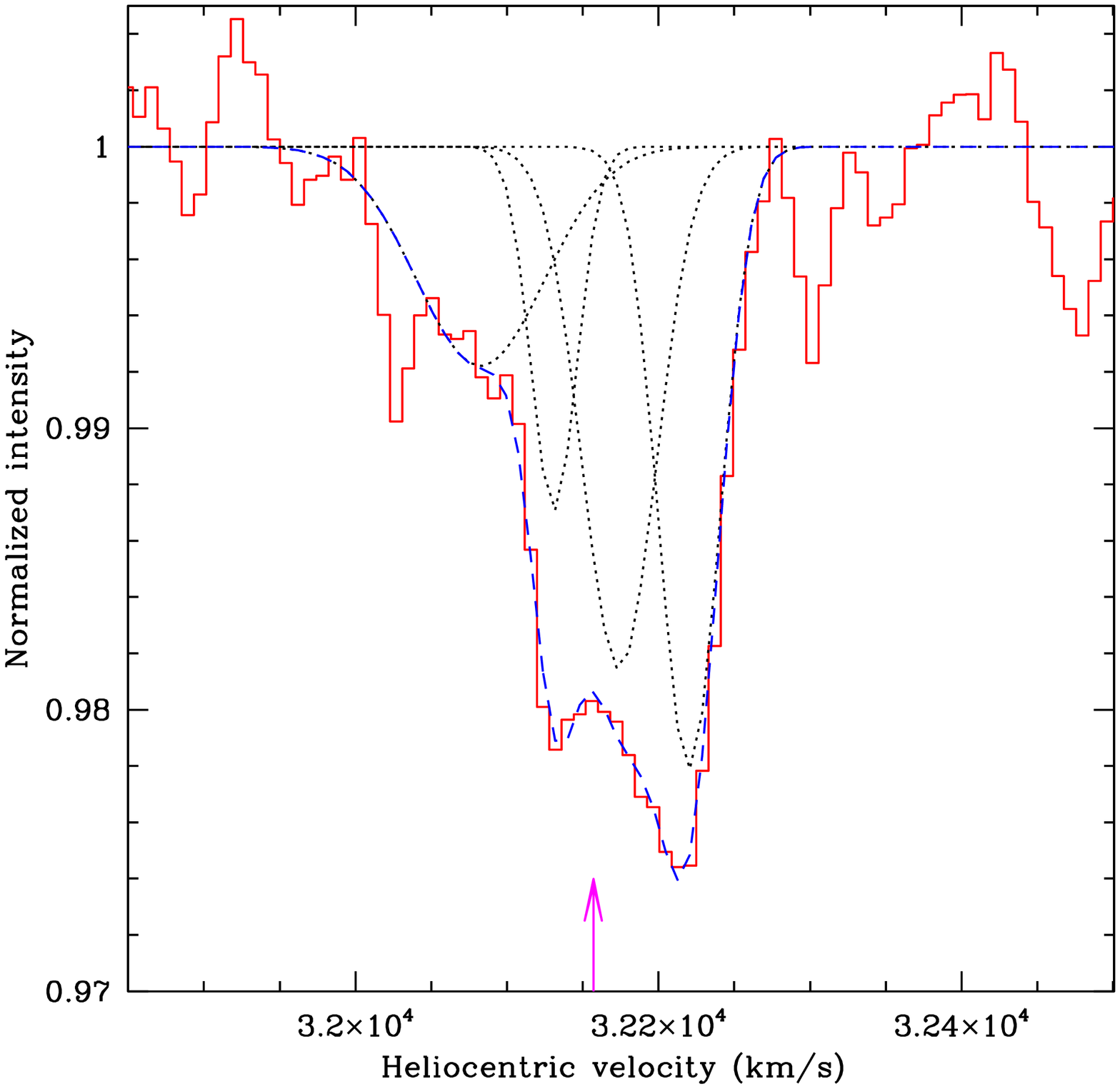}
   \includegraphics[angle=0, width=.32\textwidth,]{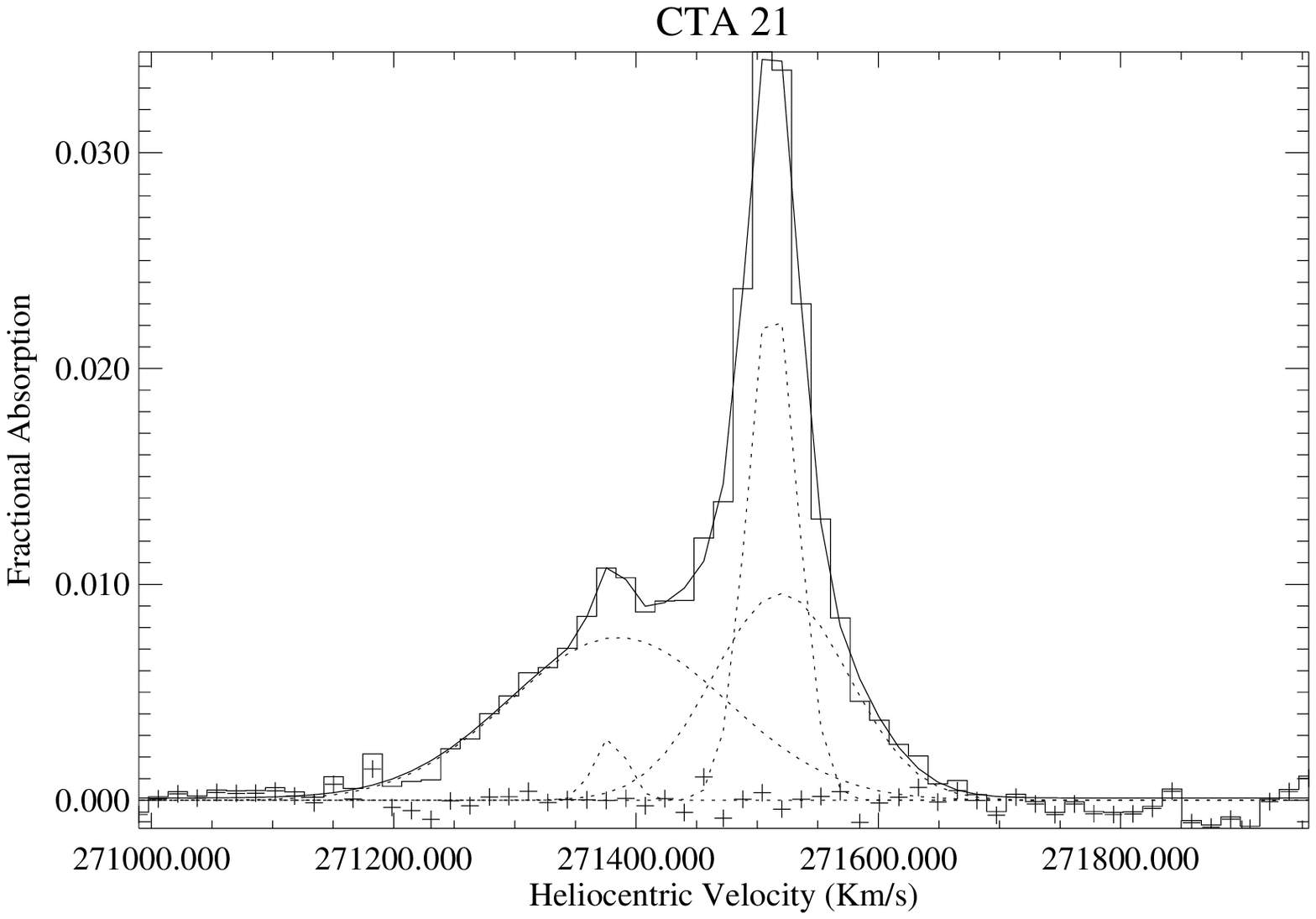}
       }
}
\caption{H{\sc i} absorption spectra towards 4C29.30 (left panel \cite{37}, J1247+6723 (middle panel \cite{31}),
both obtained with the GMRT, and towards CTA21 (right panel \cite{17}) obtained with the Arecibo telescope. 
        }
\end{center}
\end{figure}

\section{Recurrent activity and the supply of gas}
It is often speculated that the AGN activity may be triggered by
a fresh supply of gas, possibly due to interactions with a companion. 
To test this scenario, there have been a number of searches for 
molecular and atomic gas in the central regions of the rejuvenated 
radio galaxies. Saripalli \& Mack \cite{29}  observed a sample of 9 radio
galaxies with evidence of recurrent activity with the IRAM 30-m telescope
to detect CO. The only detection was 3C293, which was the only FRII radio
galaxy known to have CO in an earlier survey of IRAS flux-density limited
survey of radio galaxies \cite{30}.

Saikia, Gupta \& Konar \cite{31} reported the detection of H{\sc i} absorption towards
the inner VLBI-scale double of the DDRG J1247+6723, and noted the high incidence
of H{\sc i} absorption in sources with evidence of recurrent activity. Other 
examples include 3C236 \cite{32,33}, 
3C293 \cite{34,35}, Cyg A \cite{34}, 4C29.30 \cite {37}, Cen A \cite{21}
and CTA21 \cite{17}. Although the number of sources is small, the 
detection of H{\sc i} absorption appears to be more frequent than for compact steep-spectrum
 and GPS sources \cite{38}. The line profiles
of H{\sc i} absorption towards the cores often tend to be complex with 
multiple components. It would be interesting to observe these sources 
in H{\sc i} with VLBI-scale resolution to probe the distribution and properties
of the absorbing gas.

\section{Concluding remarks}
We have shown examples of recurrent AGN activity in radio galaxies and quasars 
inferred from radio and X-ray observations of radio sources in relatively low-density
environments to those in clusters of galaxies. In addition to radio galaxies and
quasars, it is also important to systematically probe for evidence of recurrent
activity in other classes of AGN, such as for example the Seyfert galaxies. 
The time scales of jet interruption inferred from the present studies of radio galaxies 
range from 10$^5$ to 10$^8$ yr, and models would have to explain the large range of time
scales. It is also possible that different models may be applicable to explain
episodic activity on different time scales. Early trends appear to indicate that
the radio galaxies with evidence of recurrent activity tend to exhibit H{\sc i}
in absorption with column densities in the range $\sim$8$-$50$\times$10$^{20}$ cm$^{-2}$
and often with complex line profiles. It would also be interesting to explore possible
relationships between the coupling of episodic star formation and the rejuvenation
of AGN activity, both of which may be intimately related to the supply of gas. Such
studies for the giant radio galaxy 3C236 suggest that the AGN fuel supply may have been
interrupted for $\sim$10$^7$ yr due to a minor merger event and has now been restored \cite{41}.
Deeper multi-wavelength studies
are likely to give us new insights towards understanding recurrent AGN
activity, and its effects on feedback processes and galaxy evolution.


\begin{thebibliography}{99}

\bibitem{1} Rees M.J., 1984, ARA\&A, 22, 471

\bibitem{2} Begelman M.C., Li Z.Y., 1994, ApJ. 426, 259

\bibitem{3} Sikora M., Stawarz {\L}., Lasota J.P., 2007, ApJ, 658, 815

\bibitem{4} Marconi A., Risaliti G., Gilli R., Hunt L.K., Maiolino R., Salvati M., 2004, MNRAS, 351, 169

\bibitem{5} Saikia D.J., Jamrozy M., 2009, BASI, 37, 63

\bibitem{6} Nipoti C., Blundell K.M., Binney J., 2005, MNRAS, 361, 633
   
\bibitem{7} Schoenmakers A.P., de Bruyn A.G., R\"{o}ttgering H.J.A., van der Laan H., Kaiser C.R., 2000, MNRAS, 315, 371

\bibitem{8} Subrahmanyan R., Saripalli L., Hunstead R.W., 1996, MNRAS, 279, 257

\bibitem{9} Lara L., M\'arquez I., Cotton W.D., Feretti L., Giovannini G., Marcaide J.M., Venturi T., 1999, A\&A, 348, 699

\bibitem{10} Steenbrugge K.C., Blundell K.M., Duffy P., 2008, MNRAS, 388, 1465

\bibitem{11} Steenbrugge K.C., Heywood I., Blundell K.M., 2010, MNRAS, 401, 67

\bibitem{12} Erlund M.C., Fabian A.C., Blundell K.M., Celotti A., Crawford C.S., 2006, MNRAS, 371, 29

\bibitem{13} David L.P., Jones C., Forman W., Nulsen P., Vrtilek J., O'Sullivan E., Giacintucci S., Raychaudhury S., 2009,
              ApJ, 705, 624

\bibitem{14} Giacintucci S., Vrtilek J.M., O'Sullivan E., Raychaudhury S., David L.P., Venturi T., Athreya R., Gitti M., 2009,
           in The Monster's Fiery Breath: Feedback in Galaxies, Groups, and Clusters 

\bibitem{15} Saikia D.J., Konar C., Kulkarni V.K., 2006, MNRAS, 366, 1391

\bibitem{16} Joshi S.A., Nandi S., Saikia D.J., Ishwara-Chandra C.H., Konar C., 2011, MNRAS, in press 
  
\bibitem{17} Salter C.J., Saikia D.J., Minchin R., Ghosh T., Chandola Y., 2010, ApJL, 715, 117
   
\bibitem{18} Sirothia S.K., Saikia D.J., Ishwara-Chandra C.H., Kantharia N.G., 2009, MNRAS, 392, 1403
  
\bibitem{19} Giacintucci S., Venturi T., Murgia M., Dallacasa D., Athreya R., Bardelli S., Mazzotta P., Saikia D.J., 2007, A\&A, 476, 99
   
\bibitem{20} Giacintucci S., et al., 2011, ApJS, submitted (2011arXiv1103.1364G)

\bibitem{21} Morganti R., Oosterloo T., Struve C., Saripalli L., 2008, A\&A, 485, L5 
   
\bibitem{22} Gizani N.A.B., Leahy J.P., 2003, MNRAS, 342, 39

\bibitem{23} Jamrozy M., Saikia D.J., Konar C., 2009, MNRAS, 399, L141

\bibitem{24} Kaiser C.R., Schoenmakers A.P., R\"{o}ttgering H.J.A., 2000, MNRAS, 315, 38

\bibitem{25} Jamrozy M., Konar C., Saikia D.J., Stawarz {\L}., Mack K.-H., Siemiginowska A., 2007, MNRAS, 378, 581

\bibitem{26} Konar C., Saikia D.J., Jamrozy M., Machalski J., 2006, MNRAS, 372, 693

\bibitem{27} Machalski J., Jamrozy M., Konar C., 2010, A\&A, 510, 84 

\bibitem{28} Ishwara-Chandra C.H., Saikia D.J., 1999, MNRAS, 309, 100

\bibitem{29} Saripalli L., Mack K.-H., 2007, MNRAS, 376, 1385

\bibitem{30} Evans A.S., Mazzarella J.M., Surace J.A., Frayer D.T., Iwasawa K., Sanders D.B., 2005, ApJS, 159, 197

\bibitem{31} Saikia D.J., Gupta N., Konar C., 2007, MNRAS, 375, L31

\bibitem{32} Schilizzi R.T. et al., 2001, A\&A, 368, 398

\bibitem{33} Conway J.E., Schilizzi R.T., 2000, in Conway J.E., Polatidis A.G., Booth R.S., Pihlstr\"om Y.M., eds,
              EVN Symp. 2000, Onsala Space Observatory, p. 123

\bibitem{34} Conway J.E., 1998, ASPC, 144, 231

\bibitem{35} Beswick R.J., Peck A.B., Taylor G.B., Giovannini G., 2004, MNRAS, 352, 49

\bibitem{36} Emonts B.H.C., Morganti R., Tadhunter C.N., Oosterloo T.A., Holt J., van der Hulst J.M.,
              2005, MNRAS, 362, 931

\bibitem{37} Chandola Y, Saikia D.J., Gupta N., 2010, MNRAS, 403, 269 (arXiv:0910.4427)

\bibitem{38} Gupta N., Salter C.J., Saikia D.J., Ghosh T., Jeyakumar S., 2006, MNRAS, 373, 972 

\bibitem{39} Hintzen P., Ulvestad J., Owen F., 1983, AJ, 88, 709 

\bibitem{40} Reynolds C.S., Begelman M.C., 1997, ApJ, 487, L135

\bibitem{41} Tremblay G.R., O'Dea C.P., Baum S.A., Koekemoer A.M., Sparks W.B., de Bruyn G., Schoenmakers A.P., 2010, ApJ, 715, 172

\bibitem{42} Brocksopp C., Kaiser C.R., Schoenmakers A.P., de Bruyn A.G., 2007, MNRAS, 382, 1019

\bibitem{43} Brocksopp C., Kaiser C.R., Schoenmakers A.P., de Bruyn A.G., 2011, MNRAS, 410, 484

\bibitem{44} Barthel P.D., 1989, ApJ, 336, 606

\bibitem{45} Mocz P., Fabian A.C., Blundell K.M., 2011, MNRAS, in press

\bibitem{46} Blundell K.M., Fabian A.C., 2011, MNRAS, 412, 705

\end{thebibliography}
\end{document}